\documentclass[twocolumn, prl]{revtex4}

\usepackage{graphicx}
\usepackage{dcolumn}
\usepackage{bm}

\begin{document}

\title{The behavior of the entanglement entropy in interacting quasi-1D systems and its consequences for their efficient numerical study}

\author {Samuel Moukouri and Eytan Grosfeld}

\affiliation{Department of Physics, Ben-Gurion University of the Negev, Be'er-Sheva 84105, Israel} 

\begin{abstract}
The density matrix renormalization group (DMRG) method allows an efficient computation of the properties of interacting 1D quantum systems. Two-dimensional (2D) systems, capable of displaying much richer quantum behavior, generally lie beyond its reach except for very small system sizes. Many of the physical properties of 2D systems carry into the quasi-1D case, for which, unfortunately, the standard 2D DMRG algorithm fares little better. By finding the form of the entanglement entropy in quasi-1D systems, we directly identify the reason for this failure. Using this understanding, we explain why a modified algorithm, capable of cleverly exploiting this behavior of the entanglement entropy, can accurately reach much larger system sizes. We demonstrate the power of this method by accurately finding quantum critical points in frustration induced magnetic transitions, which remain inaccessible using the standard DMRG or the Monte Carlo methods.
\end{abstract}

\maketitle

The study of ground-state phase transitions in simple quantum
models in two dimensions (2D) is essential in understanding the physics 
of models of various layered systems including the cuprate
superconductors, the organic conductors and quantum magnets. A considerable 
effort is currently devoted to the development of powerful numerical algorithms
for 2D systems \cite{schollwock,maier}. The entanglement entropy (EE), $S_E$, 
has emerged as an important quantity that can gauge the validity of these algorithms \cite{holzhey,jin,calabrese,gioev,barthel,eisert}. 
$S_E$ obeys strict area laws. In one dimension (1D), the entropy area 
law  prescribes that $S_E \propto c$ for a gapped system (where $c$ is a constant) and that $S_E \propto \log_2 L$ for a
gapless system (where $L$ is the linear dimension of the subsystem). It was
later realized that it is this mild growth of $S_E$ with the subsystem size that
lies behind the extraordinary accuracy of the density-matrix renormalization 
group (DMRG) \cite{white} method in 1D. In 2D, however,
$S_E \propto L$ for gapped systems, and $S_E \propto L \log_2 L$ for gapless
systems. This fast growth of $S_E$ with $L$ reflects itself
in the DMRG method as an exponential increase in the number of states
necessary to keep the truncation error small. As a consequence, the direct application of the conventional DMRG to 2D systems has been limited to
long systems with relatively narrow width. However, in situations where the correlation length is
larger than this width, as in the vicinity of quantum critical points, uncertainties in the extrapolated DMRG data may occur and another approach may be desirable.

Novel ideas to extend DMRG to higher dimensions involve the use of
a class of variational states, the tensor network states, which
satisfy the area law by construction. Among these states are the
projected entanglement pairs \cite{verstraete1,verstraete2} which are the generalization
of matrix product states to higher dimensions. Another type
of states are obtained by the multiscale entanglement renormalization
ansatz which consists of application of isometries and disentanglers
on tensor network states \cite{vidal1,vidal2}. These promising ideas have not, however,
shown so far any decisive superiority over the conventional 2D DMRG \cite{stoudemire}.

Another route for exploring quantum phase transitions in 2D involves the use of quasi-1D systems. In these systems the smallness of the transverse coupling may enable the use of a more traditional DMRG approach. Highly anisotropic quasi-1D systems are known to display both 1D and 2D characteristics. Coexistence of long-range
order and Haldane gap excitations has been observed in neutron scattering experiments of mixed-spin nickelates R$_2$BaNiO$_5$ where R$=$Pr, Nd, Nd$_x$Y$_{1-x}$ 
\cite{zheludev1,zheludev2}. The broad continuum of spinons in the frustrated 
anisotropic antiferromagnet Cs$_2$CuCl$_4$ \cite{coldea} was argued to consist of 
descendents of the pure 1D spinons \cite{kohno}. In quasi-1D organic
conductors (Bechgaard salts) low-energy excitations are well described
within the Fermi liquid theory whereas signatures of 1D physics are
observed in optical \cite{schwartz} and transport \cite{moser} measurements 
at higher energy. These raise an intriguing question: how does the entropy area law apply in quasi-1D systems? Does $S_E$ simply behave as in higher
dimensional systems or is the 1D nature of these systems also reflected 
in $S_E$?

In this paper, we explore the behavior of the EE in an interacting quasi-1D system using an approximate DMRG algorithm, the two-step DMRG \cite{moukouri1}. When applied to a typical interacting system, the quasi-1D Heisenberg system with $S=1$, we show that the EE satisfies the 2D area law as far as the coupling between the chains is small. Importantly, the magnitude of the EE is also decided by a small multiplicative coefficient, governed by the smallness of the transverse coupling $J_y\ll 1$. As we now argue, this particular form for the area law in quasi-1D systems supports the applicability of the algorithm in this class of systems and explains the failure of the traditional 2D DMRG in comparable system sizes. This understanding paves the way to numerical studies of quantum phase transitions in quasi-1D systems. As evidence for the power of this algorithm, we compute
the critical point in the interchain driven magnetic transition and show that its value accurately compares with quantum Monte Carlo. We then study the phase diagram in the presence
of frustration. Frustrated quantum Hamiltonians lie beyond the reach of the quantum Monte Carlo method which is
hampered by the minus sign problem. Remarkably, we find that the maximally frustrated point in the spin-Peierls phase corresponds to minimal EE. This means paradoxically that the two-step DMRG performs better in the regime of strong frustration.

\begin{figure}[b]
\begin{center}
$\begin{array}{c@{\hspace{0.25in}}c}
\vspace{0.25in}
\includegraphics[width=2.75cm, height=2.5cm]{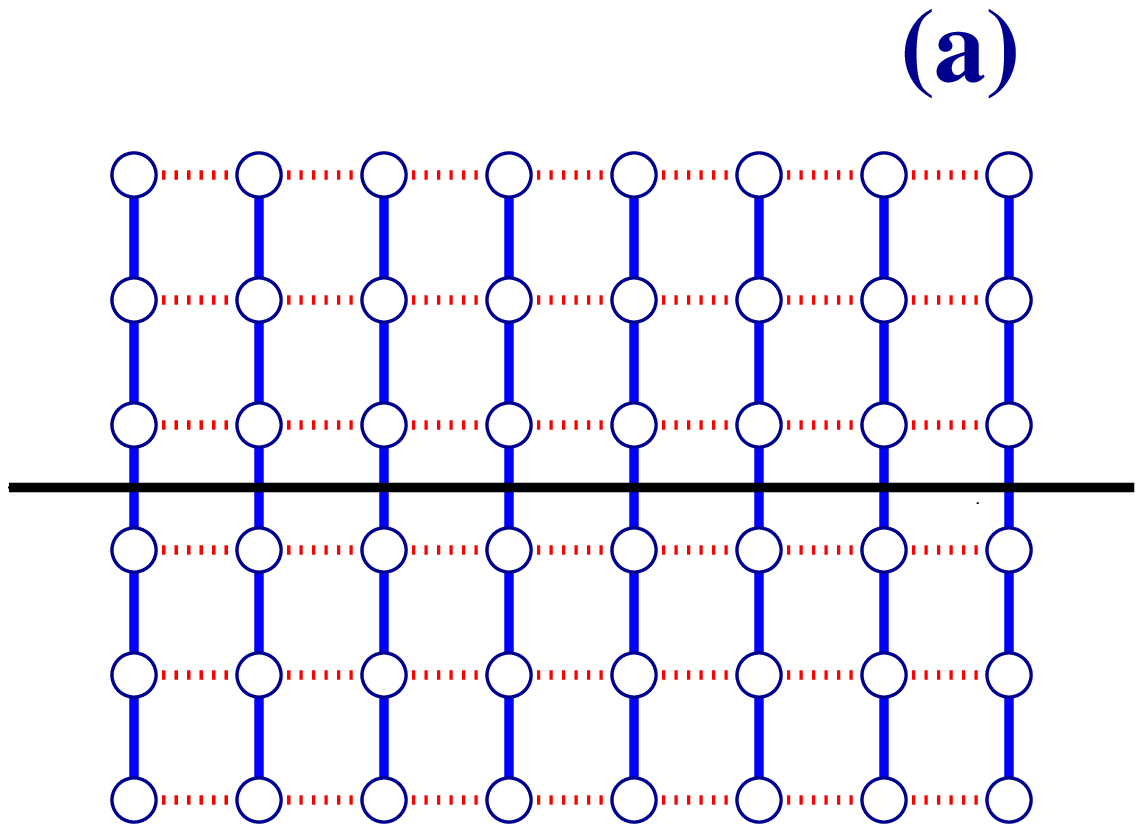}
\hspace{0.25in}
\includegraphics[width=2.75cm, height=2.5cm]{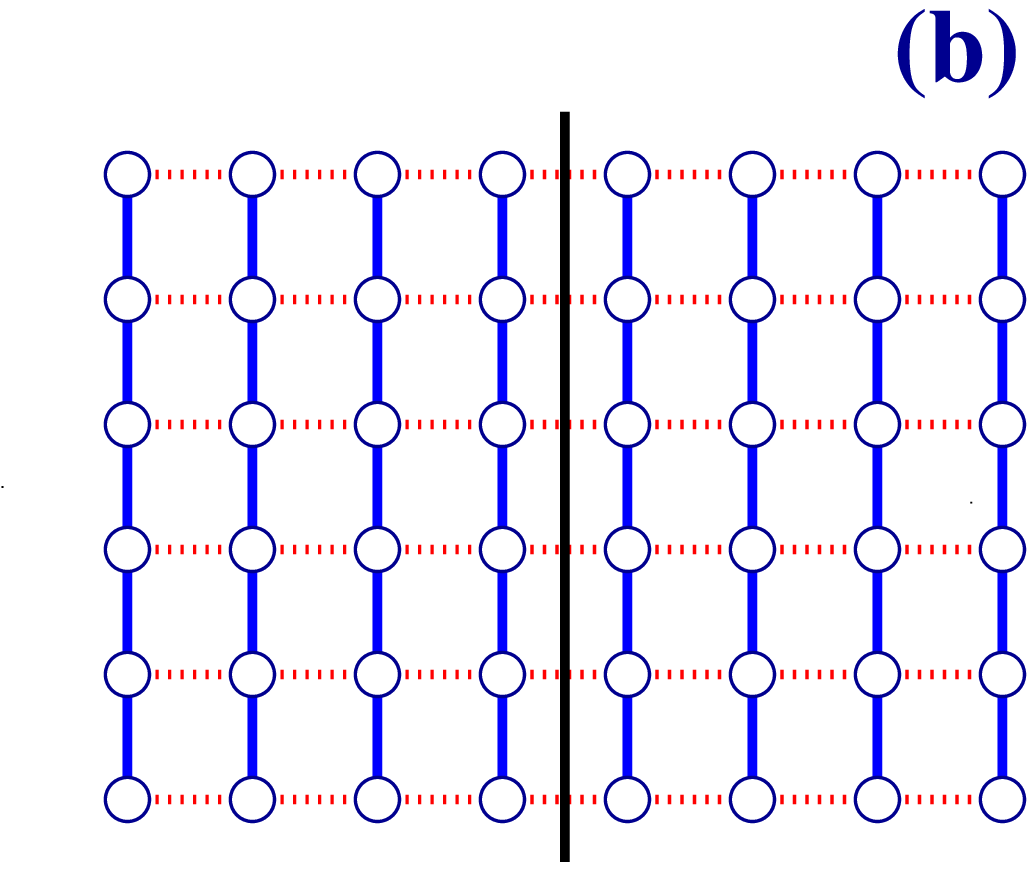}
\end{array}$

$\begin{array}{c@{\hspace{0.25in}}c}

\includegraphics[width=2.5cm, height=2.5cm]{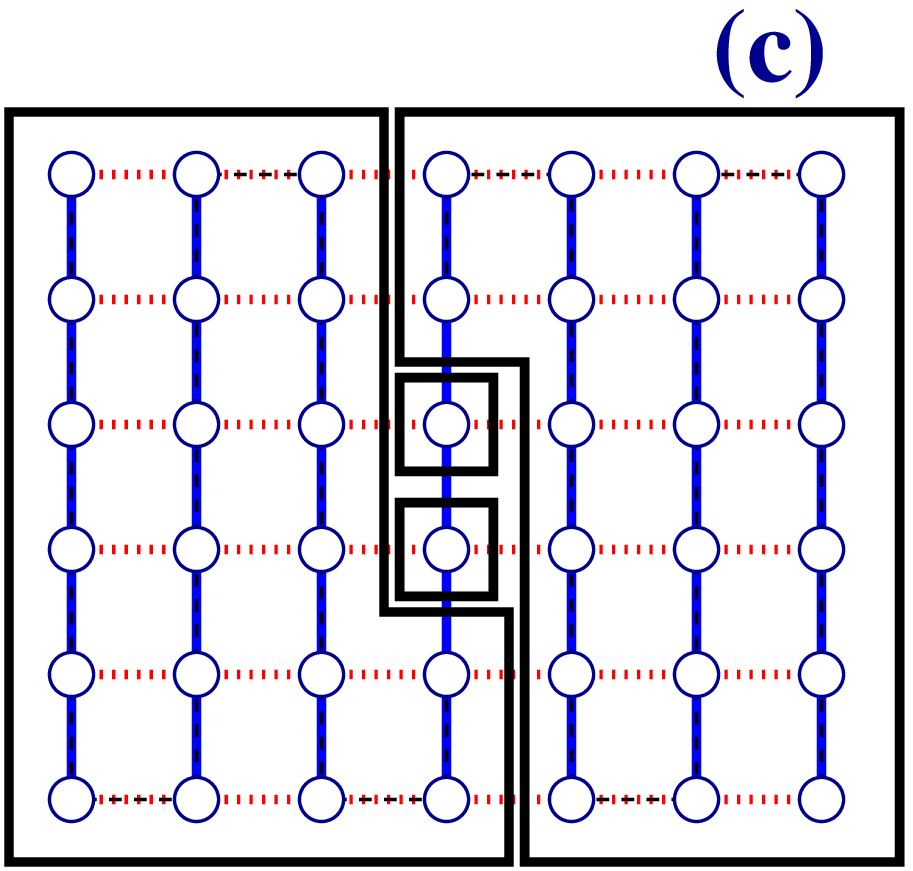}
\hspace{0.25in}
\includegraphics[width=2.5cm, height=2.5cm]{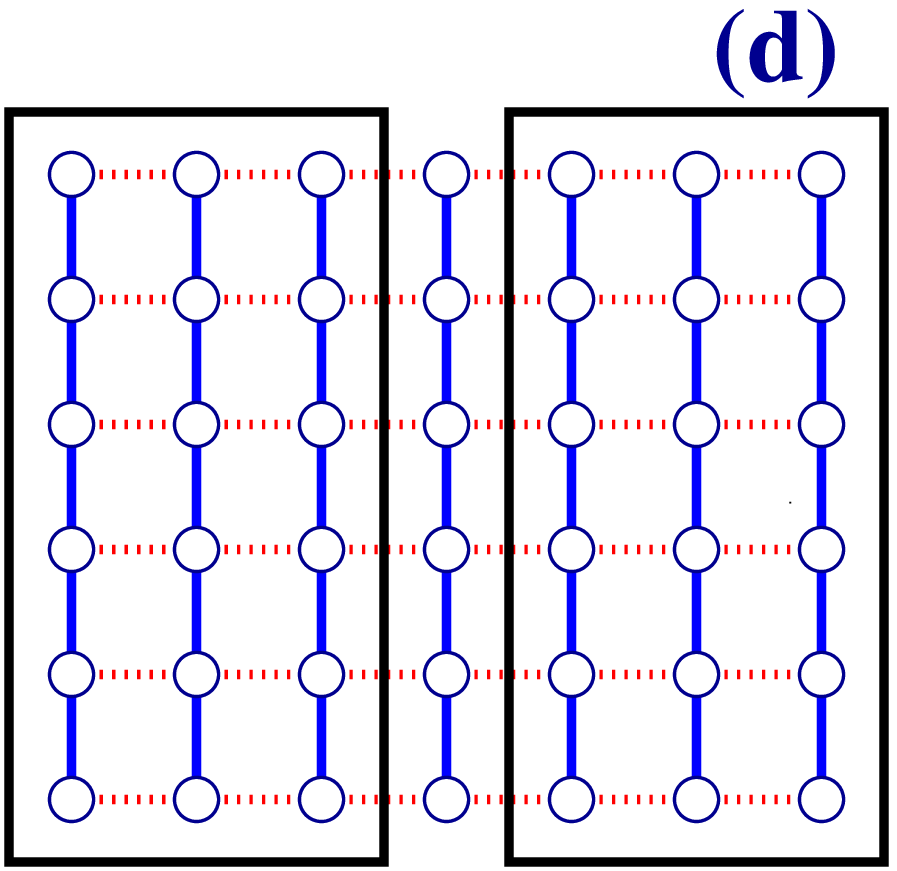}
\end{array}$
\end{center}
\caption{(a) Horizontal and (b) vertical partitions of the quasi-1D
lattice; (c) generation of the quasi-1D lattice from conventional
DMRG: the system is grown in the $x$ and $y$ direction at the
same time by injection of new sites in the middle; (d) from the 
two-step DMRG which exploits the anisotropy of the system: a chain of 
$L_x$ is first built in the first step then the system is grown in the 
$y$ direction for a fixed $L_x$. Here $J_x$ coupling is denoted by a blue solid line and $J_y$ coupling by a dashed red line.}
\label{clusters}
\end{figure}

In constructing algorithms for highly anisotropic quasi-1D systems, it is 
crucial to take into account the effect of the strong anisotropy in order
to avoid the high cost in EE that would come from a
full 2D treatment. To illustrate this point, let us consider the following 
Heisenberg model with $S=1$ in an anisotropic $L_x \times L_y$
lattice,
\begin{eqnarray}
 \nonumber  H_s= J_x\sum_{i_x,i_y} {\bf S}_{i_x,i_y}{\bf S}_{i_x+1,i_y}+
       J_y\sum_{i_x,i_y} {\bf S}_{i_x,i_y}{\bf S}_{i_x,i_y+1}+\\
       J_d\sum_{i_x,i_y} ({\bf S}_{i_x,i_y}{\bf S}_{i_x+1,i_y+1}+
                           {\bf S}_{i_x+1,i_y}{\bf S}_{i_x,i_y+1}),
\label{hamiltonian}
\end{eqnarray}
where $i_x=1,...,L_x$, $i_y=1,...,L_y$, stand respectively for the 
chain and the inter-chain indices. Here $J_x$ is the nearest-neighbor exchange
along the chains; $J_y$ and $J_d$ are respectively the nearest-neighbor and
the next-nearest neighbor exchange parameters between the chains. The 
system is taken to be highly anisotropic, $J_y,J_d \ll J_x=1$. As we now explain, it may
therefore prove beneficial to study it in two DMRG steps.

The two-step DMRG is an adaptation of conventional functional renormalization approaches to quasi-1D 
systems. When the transverse coupling is far smaller than the intra-chain coupling, 
the functional integral approach is based on a hierarchy of the energy scales 
(see for instance Ref.~\onlinecite{bourbonnais-caron}), 
and involves the use of the disconnected chain ($J_y=J_d=0$) as the starting point. 
This may be justified by the observation of 1D physics at high energy or temperature in 
quasi-1D systems. 
When translated to ground-state DMRG computations, use of the disconnected chain basis as a starting point will be justified if
$J_y,J_d \ll \Delta E(L_x)$. Here $\Delta E(L_x)$ is the energy width of
the states which are retained to describe each chain. In the two-step
DMRG, the first term of $H_s$, $ h_\ell=\sum_{i_x} {\bf S}_{i_x,i_\ell}{\bf S}_{i_x+1,i_\ell}$, 
is solved for each chain $\ell$ using the 1D DMRG. Then $H_s$ is projected 
on the direct product basis of the low-lying states of the $h_\ell$ chains. 
This yields an effective 1D Hamiltonian ${\tilde H}_s$ which is solved using 
the 1D DMRG,
\begin{eqnarray}
            {\tilde H}_s= \sum_\ell h_\ell  +
       J_y\sum_{\ell} {\bf {\tilde S}}_\ell \cdot {\bf {\tilde S}}_{\ell+1}+
       J_d\sum_{\ell} {\bf {\tilde S}}_{\ell} \odot {\bf {\tilde S}}_{\ell+1}.              
\label{rhamiltonian}
\end{eqnarray}
Here ${\bf {\tilde S}}_\ell$ are block spin vectors representing all
the spins in a chain $\ell$. The operators acting between them are defined as in $H_s$ in Eq.~(\ref{hamiltonian}). 
This procedure may be viewed as inverting the mapping used in Ref.~\onlinecite{schulz}, where a 1D large-spin $S$ system was analyzed as $2S$ coupled spin-$\frac{1}{2}$ chains.

\begin{figure}[t]
\begin{center}
$\begin{array}{c@{\hspace{0.5in}}c}
\includegraphics[width=6.cm, height=4.cm]{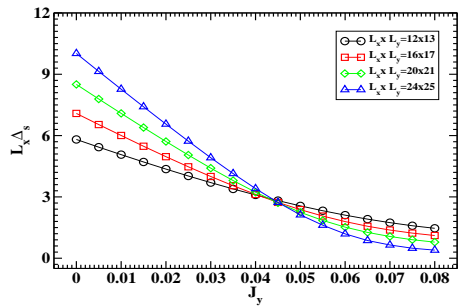}
\end{array}$
\end{center}
\caption{$\Delta_s \times L_x$ as function of $J_y$ for coupled
  $S=1$ chains.}
\label{gaps}
\end{figure}

Setting $J_d=0$ in Eq.~(\ref{hamiltonian}), in Ref.~\onlinecite{moukouri3} we studied systems of 
up to $24 \times 25$ spins. The two-step DMRG results were found to be in 
excellent agreement with quantum Monte Carlo simulations \cite{matsumoto}
which are known to be very accurate for unfrustrated quantum spin systems.
For completeness, we reproduce in Fig.~\ref{gaps} an improved finite size 
scaling of the spin gap $L_x \Delta_s(L_x\times L_y)$ as function of $J_y$.
In this calculation, we kept $ms_1=729$ and $ms_2=135$ states 
respectively in the first and second steps of the DMRG.
There is a ground-state phase transition driven by the transverse 
coupling $J_y$ between a gapped phase at $J_y < J_{y_c}$ and a gapless 
phase with magnetic long-range order for $J_y > J_{y_c}$. The curves converge at the critical 
point since, using the relevant scaling function, $ L_x\Delta_s(L_x\times L_y)=f\left(C(J_y-J_{y_c})L_x^{1/\nu}\right)$ is
independent of the size at $J_y=J_{y_c}$. 

Fig.~\ref{gaps} displays the finite size behavior of $\Delta_s$. The data 
converges around $J_{y_c} \approx 0.045$. In Ref.~\onlinecite{moukouri3}, where 
a more detailed analysis was made in the vicinity of the critical point, the 
two-step DMRG predicted the value $J_{y_c}=0.04367$ \cite{moukouri3}, 
in excellent agreement with quantum Monte Carlo result, 
$J_{y_c}=0.043648 (8) J_x$ \cite{matsumoto}.  
For the largest system studied, $L_x \times L_y=24 \times 25$, a typical 
truncation error is $\rho_1=1. \times 10^{-7}$ during the first DMRG step.
We targeted the lowest states in seven spin sectors having total $z$-component
of the spin $S_z=0$, $\pm 1$, $\pm 2$, and $\pm 3$. For the largest size
studied, the lowest states of $S_z=\pm 4$ are higher than the highest states
kept in the $S_z=\pm3$ sectors, hence, they were  not retained. In the 
second step when two states are targeted, $\rho_2= 1. \times 10^{-6}$ at 
$J_y=0.02$ in the disordered phase, $\rho_2=3.25 \times 10^{-5}$ at $J_y=0.045$ 
in the vicinity of the quantum critical point, and $\rho_2=1.94 \times 10^{-4}$ 
in the magnetically ordered phase. The energy width of the $ms_2$ states
retained was $\Delta E=2.09$. The condition of validity of the two-step
DMRG $J_y \ll \Delta E$ is thus fulfilled for all $J_y \alt 0.08$ which
were studied. The increase in $\rho_2$ may be explained as follows. 
When $J_y \ll J_{y_c}$ the
spectrum of the reduced density matrix is dominated by a single state,
namely the direct product of the wave function of the individual chains.
As $J_y$ is increased towards the magnetically ordered phase, more and 
more states gain weight in the spectrum of the reduced density matrix, increasing the truncation error.

Given the constraint imposed by the entropy area law for 2D systems, it may be
surprising that the two-step DMRG can reach large system sizes (say of the order of $24\times 25$ spins discussed above), 
yielding an excellent estimate for the quantum critical point. For comparison, we note that in 
Ref.~\onlinecite{jiang}, where the conventional DMRG was applied, only 
systems of up to $L_x \times L_y=8 \times 8$ could be reached, which is 
not enough to perform finite size analysis in the vicinity of the quantum 
critical point. However, as we now show, our results are fully consistent with entropy 
growth in 2D systems. 
First let us give an intuitive reason why large systems may be reached in this case
even when the entropy grows linearly: Consider a 2D system for
which $S_{E_{2D}} \approx cL$ and, as a reference, a 1D critical chain 
for which $S_{E_{1D}} =\frac{1}{3} \log_2(L)$ \cite{jin,calabrese}; If $c$ is 
small, say $c=0.01$, $S_{E_{2D}} < S_{E_{1D}}$ as far as $L \alt 100$.

The situation discussed in the previous paragraph naturally occurs in quasi-1D systems. In the limit $J_y=J_d=0$, 
the ground-state is trivially the direct product of the lowest states of the 
chains along the $x$ direction. Each chain has 
a gap, the Haldane gap, $\Delta_1=0.4105 J_x$ \cite{golinelli}. If a cut is made along the $y$ direction, 
as illustrated in Fig.~\ref{clusters}(a), it would be expected that
the EE of the ensemble of $L_y$ chains is trivially the 
sum of the EEs on the individual chains $h_\ell$,
$S_{E_y}(\otimes_{\ell=1}^{L_y} h_\ell)=\sum_{\ell=1}^{L_y} S_E(h_\ell)$. Given that 
$S_E(h_\ell)\approx c_H$, where $c_H\sim 1$ for each Haldane chain, we expect that $S_{E_y} \approx L_y$. 
This trivial entanglement is present despite the chains being disconnected. 
For a cut along the $x$ direction as illustrated in Fig.~\ref{clusters}(b), 
since the chains are disconnected, 
$S_{E_x}=0$. When $J_y \ne 0$ or $J_d \ne 0$, as long as $J_y,J_d \ll J_x$, 
it is expected that $S_{E_x} \ll S_{E_y}$.

The conventional DMRG, when applied to a highly anisotropic system, 
builds the quasi-1D system by growing the lattice simultaneously in the 
$x$ and $y$ directions, as illustrated in Fig.~\ref{clusters}(c). 
It is therefore expected that it will run into difficulties imposed by
the linear growth in the trivial entanglement $S_{E_y}$. 
In contrast, the two-step DMRG of Ref.~\cite{moukouri1} relies 
on the separation of the energy scales between the longitudinal and 
transverse directions: the chains are first built along the $x$-direction hence
$S_{E_y}$ is exactly taken into account by construction. 
One may worry that since the starting point for the two-step DMRG involves the disconnected chains, once $J_y\neq 0$,
the algorithm struggles with a growth of entanglement that scales with the linear size of the system. By taking this route, is crucial 
information on the EE in the quasi-1D system consequently irreversibly lost? 
And, if so, how would that reconcile with the accurate critical analysis of $\Delta_s$?

\begin{figure}[t]
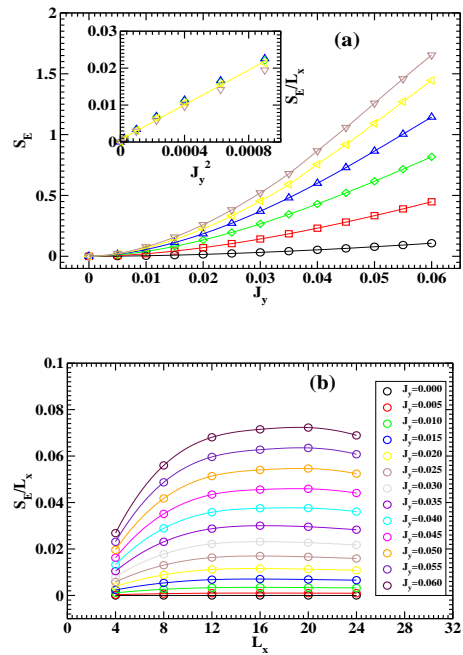

\begin{center}
$\begin{array}{c@{\hspace{0.5in}}c}
\vspace{.5cm}
\includegraphics[width=6.cm, height=4.cm]{entro1_unfrus_ins.eps}
\end{array}$
$\begin{array}{c@{\hspace{0.5in}}c}
\includegraphics[width=6.cm, height=4.cm]{entro2_unfrus_3.eps}
\end{array}$
\end{center}
\caption{(a) The EE $S_{E_x}$ in $S=1$ quasi-1D systems as 
function of $J_y$; $L_x \times L_y=4 \times 5$ (circles), $8\times 9$ 
(squares), $12 \times 13$ (diamonds), $16 \times 17$ (triangles up), 
$20 \times 21$ (triangles left), $24 \times 25$ (triangles down). 
In the inset, $S_{E_x}/L_x$ as function of $J_y^2$ in the disordered
phase, $0 \alt J_y \alt 0.03$. Data for $L_x=4$ and $L_x=8$ which have
strong size dependance were omitted. (b) $S_{E_x}/L_x$ as function of 
$L_x$ for $J_y=0$ to $J_y=0.06$.}
\label{entro-unfrus}
\end{figure}

In Fig.~\ref{entro-unfrus}, $S_{E_x}$ is shown for values of $J_y$ accross 
the quantum critical point, $J_y=0$ to $J_y=0.06 J_x$, for systems ranging 
from $4 \times 5$ to $24 \times 25$.  The $4 \times 5$ systems are essentially
exact because all the states were kept after the first step and only 
one DMRG iteration was performed during the second step. 
$S_{E_x}$ was obtained from the spectrum of the reduced density matrix
after a cut in the $x$ direction, $S_{E_x}=-\sum_{i}\lambda_i\log_2(\lambda_i)$,
where the $\lambda_i$ are the reduced density matrix eigenvalues. 
$S_{E_x}$ is expected to follow the 2D area law $S_{E_x} \propto c(J_y) L_x$, and 
$S_{E_x} \propto c(J_y) L_x \log_2(L_x)$  respectively in the gapped and the
gapless phases. In the limit $J_y=0$ we have $c(J_y)=0$; hence, from continuity we
expect $c(J_y) \ll 1$ if $J_y \ll 1$.  This small value of $c(J_y)$ is
the source of the slow growth of $S_{E_x}$ although its depends linearly on 
$L_x$.  In Fig.~\ref{entro-unfrus}(a), $S_{E_x}$ is shown as function of
$J_y$. Clearly $S_{E_x}$ remains small for very small values of $J_y$. Then, except
for very small sizes, it grows rapidly as the magnetically ordered phase is 
approached. In the inset, $S_{E_x}/L_x$ is displayed as function of $J_y^2$
in the disordered phase, far from the quantum critical point. The data
suggests that $c(J_y) \propto J_y^2$. In Fig.~\ref{entro-unfrus}(b), 
$S_{E_x}/L_x$ is plotted against $L_x$ for $J_y$ ranging from $0$ to $0.06$. 
The data is consistent with the 2D entropy area law when $J_y \alt J_{y_c}$ in the spin gap phase.
As $J_y$ is increased in the magnetically ordered phase, $S_{E_x}/L_x$ does
not show a plateau. It instead increases slowly with $L_x$. This is 
consistent with its expected logarithmic growth in this phase. 
However, $S_{E_x}/L_x$ displays a mild decay from $L_x=20$ to $L_x=24$. 
This deviation from the area law correlates with the increase of the 
truncation error seen in the calculation of $\Delta_s$.

\begin{figure}[t]
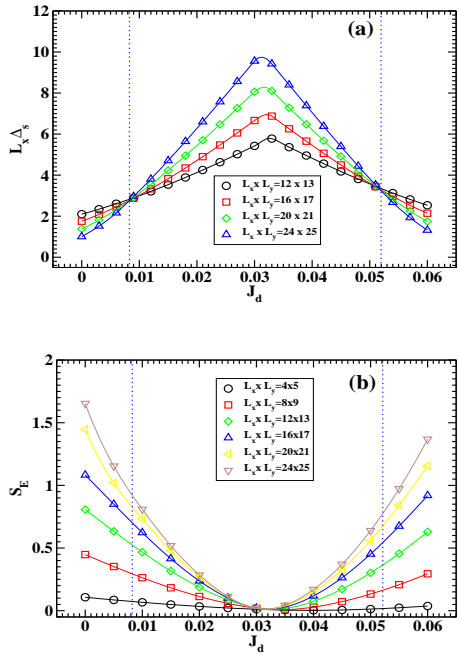

\begin{center}
$\begin{array}{c@{\hspace{0.5in}}c}
\vspace{.5cm}
\includegraphics[width=6.cm, height=4.cm]{gapgp0.eps}
\end{array}$
$\begin{array}{c@{\hspace{0.5in}}c}
\includegraphics[width=6.cm, height=4.cm]{entro1_frus.eps}
\end{array}$
\end{center}
\caption{ (a) $\Delta_s \times L_x$ as function of $J_d$ for coupled
  $S=1$ chains at $J_y=0.06J_x$. The two vertical blue dotted lines are
in the location of the quantum critical points $J_{d_{c1}}$ and $J_{d_{c2}}$.
(b) $S_{E_x}$ as function of $J_d$ for systems ranging from $4 \times 4$
to $24 \times 24$.}
\label{frus-chains}
\end{figure}

When frustration is added, $J_d\neq 0$, there is good numerical evidence
from the two-step DMRG \cite{moukouri2} and the conventional DMRG \cite{jiang} that the ground state phase diagram displays three
phases: a magnetically ordered phase with $q=(\pi,\pi)$, a
disordered phase with a spin gap, and a second magnetically
ordered phase with $q=(\pi,0)$. The signatures of these three
phases can be seen in the finite size analysis of the spin
gap shown in Fig.~\ref{frus-chains}(a). We set $J_y=0.06$,
hence at $J_d=0$, the system is in the magnetic $q=(\pi,\pi)$
phase where $L_x \Delta_s$ is a decreasing function of $L_x$.
As $J_d$ increases, the system reaches a first quantum
critical point $J_{d_{c1}}\approx 0.008$. It enters a spin gap
phase with short-range antiferromagnetic inter-chain correlations.
In the disordered phase, $L_x \Delta_s$ increases with $L_x$.
The gap increases from $J_{d_{c1}}$ until it reaches a maximum at
$J_d=J_{d_{\rm{max}}} \approx 0.03$. This is the point of maximal frustration. 
From $J_{d_{\rm{max}}}$, the gap decreases and the short-range inter-chain
correlations turn ferromagnetic. When $J_d$ reaches $J_{d_{c2}} \approx 0.052$,
the systems enters the magnetic $q=(\pi,0)$ phase in which 
$L_x \Delta_s$ is a decreasing function of $L_x$. It should be
noted that the effective transverse coupling driving these transitions
is $J_{y_{\rm{eff}}}= J_y-2J_d$. $J_{d_{c1}}$ corresponds to $J_{y_{\rm{eff}}} \approx 0.044$
which coincides with $J_{y_c}$ found above in the case $J_d=0$, and $J_{d_{c2}}$
corresponds to $J_{y_{\rm{eff}}}=-J_{y_c}$.

The EE $S_{E_x}$ as function of $J_d$ is shown in Fig.~\ref{frus-chains}(b). At 
a fixed $J_d$ in the magnetic phases, $J_d \alt J_{d_{c1}}$, and 
$J_d \agt J_{d_{c2}}$, $S_{E_x}$ shows a rapid increase. In the disordered
phase $J_{d_{c1}} \alt J_d \alt J_{d_{c2}}$, there is a significantly mild
size dependence. In particular at $J_d =0.03$, for which $J_{y_{\rm{eff}}}=0$,
$S_{E_x} \approx 0$ for any $L_x$. $J_{y_{\rm{eff}}}=0$ in the current model 
is the equivalent of the Majumdar-Gosh point of the
frustrated $S=1/2$ Heisenberg chain \cite{majumdar1,majumdar2}. This point of minimal
$S_{E_x}$ corresponds to the point of maximal frustration. Since
$S_{E_x} \approx 0$ the ground state of the systems is that of nearly
independant chains.  

To conclude, our results shed light on the growth of EE in quasi-1D systems and its consequences for efficient DMRG studies of such systems. 
We demonstrated that the two-step DMRG algorithm, designed for the study of quantum phase transitions in highly anisotropic 2D Hamiltonians, satisfies the entropy area law. 
By applying this algorithm to quasi-1D systems and analyzing the results, we argued that system sizes large enough to allow accurate critical analysis can be reached, extending beyond the reach of conventional DMRG. 
The underlying reason for this increase in efficiency is that the linear growth of the EE is crucially controlled by a small transverse coupling.  
In addition, we find that frustration can generate effective small transverse couplings in spin-Peierls phases. 
In these phases, the EE reaches a minimum at the maximally frustrated point. 
This suggests that spin-Peierls wave functions with their minimal EE may be used as a variational starting point in studying 
quantum phase transitions in frustrated spin systems in the vicinity of critical points. 


\begin{acknowledgments}

We acknowledge support from the Israel Science Foundation (Grant No.~401/12) and the European Union's Seventh Framework Programme (FP7/2007-2013) under Grant No.~303742.

\end{acknowledgments}

\end{document}